RESEARCH ARTICLE

# Efficient and Anonymous Two-Factor User Authentication in Wireless Sensor Networks: Achieving User Anonymity with Lightweight Sensor Computation

Junghyun Nam[1], Kim-Kwang Raymond Choo[2], Sangchul Han[1]*, Moonseong Kim[3], Juryon Paik[4], Dongho Won[4]

1 Department of Computer Engineering, Konkuk University, Chungju, Chungcheongbukdo, Korea, 2 School of Information Technology and Mathematical Sciences, University of South Australia, Mawson Lakes, South Australia, Australia, 3 Information Management Division, Korean Intellectual Property Office, Daejeon, Korea, 4 Department of Computer Engineering, Sungkyunkwan University, Suwon, Gyeonggido, Korea

* schan@kku.ac.kr

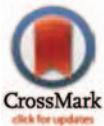

**Data Availability Statement:** All relevant data are within the paper and its Supporting Information files.

**Funding:** This work was supported by Konkuk University. The funders had no role in study design, data collection and analysis, decision to publish, or preparation of the manuscript.

**Competing Interests:** The authors have declared that no competing interests exist.

## Abstract

A smart-card-based user authentication scheme for wireless sensor networks (hereafter referred to as a SCA-WSN scheme) is designed to ensure that only users who possess both a smart card and the corresponding password are allowed to gain access to sensor data and their transmissions. Despite many research efforts in recent years, it remains a challenging task to design an efficient SCA-WSN scheme that achieves user anonymity. The majority of published SCA-WSN schemes use only lightweight cryptographic techniques (rather than public-key cryptographic techniques) for the sake of efficiency, and have been demonstrated to suffer from the inability to provide user anonymity. Some schemes employ elliptic curve cryptography for better security but require sensors with strict resource constraints to perform computationally expensive scalar-point multiplications; despite the increased computational requirements, these schemes do not provide user anonymity. In this paper, we present a new SCA-WSN scheme that not only achieves user anonymity but also is efficient in terms of the computation loads for sensors. Our scheme employs elliptic curve cryptography but restricts its use only to anonymous user-to-gateway authentication, thereby allowing sensors to perform only lightweight cryptographic operations. Our scheme also enjoys provable security in a formal model extended from the widely accepted Bellare-Pointcheval-Rogaway (2000) model to capture the user anonymity property and various SCA-WSN specific attacks (e.g., stolen smart card attacks, node capture attacks, privileged insider attacks, and stolen verifier attacks).

## Introduction

The quest to understand real-world phenomena at a fine spatial-temporal resolution has led to a great increase in the interest in wireless sensor networks (WSNs). Where not already in place, a WSN is now being planned and deployed in various application settings such as wildlife monitoring, military surveillance, healthcare diagnostics, and vehicular tracking [1]. Providing an application service in a WSN environment introduces significant security challenges for the involved parties: sensors, users and gateways. One fundamental challenge is to establish a shared session key between a sensor and a user in an authenticated manner (known as *authenticated key exchange*) via a gateway, and thereby to prevent unauthorized access to sensitive sensor data and their transmissions. Since sensors have severe resource constraints and due to network characteristics such as unattended operation and unreliable communication channel, authenticated key exchange in WSNs is generally regarded as more challenging to achieve than in traditional networks with sufficient computing resources and pre-existing infrastructures. Achieving authenticated key exchange becomes even more difficult when user anonymity is desired. As the concern for privacy increases in our lives, user anonymity has become a vital security property in various WSN applications as well as in many other applications like location-based services, e-voting, mobile roaming services, and anonymous web browsing.

A smart-card-based user authentication scheme for WSNs (in short, a SCA-WSN scheme) allows a user holding its smart card issued by the gateway to achieve authenticated key exchange with a sensor, preferably in a way that its anonymity is preserved. Since the early work of Das [2], He et al. [3], Khan and Alghathbar [4] and Chen and Shih [5], all of which provide no key-exchange functionality, the design of SCA-WSN schemes has attracted much attention from researchers due to their potential to be widely deployed, and a number of proposals offering various levels of security and efficiency have been presented [6–20]. Some schemes consider only authenticated key exchange [6, 8, 9, 12, 20] while others attempt to additionally provide user anonymity [7, 10, 11, 13–19]. Schemes such as the ones in [6, 12, 20] employ elliptic curve cryptography to provide perfect forward secrecy while most schemes [7–11, 13–19] use only lightweight cryptographic techniques, such as symmetric encryptions, message authentication codes and hash functions, to focus on improving the efficiency.

One common security requirement for SCA-WSN schemes is to ensure that:

> *only a user who is in possession of both a smart card and the corresponding password can be successfully authenticated (by the gateway) and access the sensor data.*

This requirement is commonly referred to as *two-factor security* [21–25] and is modelled via an adversary who is able to either extract all the information inside the smart card of a user or learn the password of the user, but not both. (Clearly, there is no means to prevent the adversary from impersonating a user if both the information in the smart card and the password of the user are disclosed.) The former requires physical access to the smart card and then mounting a side-channel attack [26, 27] on the (lost, misplaced or stolen) card, while the latter can be achieved with shoulder-surfing or by using a malicious card reader. Any attack exploiting the former ability is commonly called a *stolen smart card attack* and is considered practical under the assumption that users' smart cards are non-tamper-resistant. Accordingly, SCA-WSN schemes should be designed to achieve their intended security properties, such as authenticated key exchange and user anonymity, against stolen smart card attacks.

Despite the many research efforts to date, it remains a challenging task to design an efficient SCA-WSN scheme that provides user anonymity. The recent work of Wang and Wang [28, 29] shows that, under the non-tamper-resistance assumption of smart cards, no SCA-WSN scheme can provide user anonymity without recourse to public key cryptography. This result is somewhat surprising because it implies that all existing anonymous schemes using only lightweight cryptographic techniques [7, 10, 11, 13–19] fail to achieve user anonymity in the presence of an adversary who can mount a stolen smart card attack. As an example of such a failure, we here take the recent SCA-WSN scheme of Jiang et al. [19] which has been presented with a claim of user anonymity. To illustrate the failure, we only need to examine the user registration and login request phases of the scheme. Let $MK$ be the master key of the gateway $GW$, and $H$ be a cryptographic hash function. Then, the two phases proceed as follow:

**User Registration**. A user $U$ registers with $GW$ as follows:

1. $U$ chooses its identity $ID_U$ and password $PW_U$, generates a random number $r$, computes $RPW_U = H(r\|PW_U)$, and submits $ID_U$ and $RPW_U$ to $GW$ via a secure channel.

2. If $ID_U$ is valid, $GW$ generates a temporary identity for $U$, $TID_U$, and computes $TC_U = H(MK\|ID_U\|TE_U)$ and $PTC_U = TC_U \oplus RPW_U$, where $TE_U$ is the expiration time of $TID_U$. $GW$ then stores $(TID_U, ID_U, TE_U)$ in its verification table, and issues $U$ a smart card containing $\{H(\cdot), TID_U, TE_U, PTC_U\}$.

3. $U$ stores the random number $r$ into the smart card, which then holds $\{H(\cdot), TID_U, TE_U, PTC_U, r\}$.

**Login Request**. $U$ inserts its smart card into a card reader, and inputs $ID_U$ and $PW_U$. The smart card retrieves the current timestamp $T_U$, selects a random key $K_U$, and computes $TC_U = PTC_U \oplus H(r\|PW_U)$, $PKS_U = K_U \oplus H(TC_U\|T_U)$ and $C_U = H(ID_U\|K_U\|TC_U\|T_U)$. Then, $U$ sends the login request message $M_U = \langle TID_U, C_U, PKS_U, T_U \rangle$ to $GW$.

Assume an attacker $A$ who has obtained the information $\{H(\cdot), TID_U, TE_U, PTC_U, r\}$ stored on the smart card of user $U$. $A$ eavesdrops and obtains the login request message $M_U = \langle TID_U, C_U, PKS_U, T_U \rangle$, and mounts the following offline dictionary attack.

**Step 1**. $A$ makes a guess $PW'_U$ on the password $PW_U$ and computes $TC'_U = PTC_U \oplus H(r\|PW'_U)$ and $K'_U = PKS_U \oplus H(TC'_U\|T_U)$.

**Step 2**. For each possible identity $ID'_U$, $A$ computes $C'_U = H(ID'_U\|K'_U\|TC'_U\|T_U)$ and verifies the correctness of $PW'_U$ and $ID'_U$ by checking that $C'_U$ is equal to $C_U$. Note that, with an overwhelming probability, $C'_U = C_U$ if and only if $PW'_U = PW_U$ and $ID'_U = ID_U$.

**Step 3**. $A$ repeats Steps 1 and 2 until the correct password and identity are found.

This dictionary attack works because the identity space is very limited in practice, being usually even smaller than the password space [28, 29]. All other schemes using only lightweight cryptographic techniques are also vulnerable to similar dictionary attacks, as shown in [28, 29]. Note that simply using a symmetric encryption scheme cannot overcome the inherent failure. Although there are some published schemes that employ elliptic curve cryptography [6, 12, 20], these schemes were designed with no user anonymity in the first place and moreover, are not efficient in the sense that they impose expensive scalar-point multiplications on resource-constrained sensors.

In this paper, we present an efficient and provably-anonymous SCA-WSN scheme that requires sensors to perform only lightweight cryptographic operations. Our scheme employs elliptic curve cryptography but restricts its use to anonymous user-to-gateway authentication in order not to impose any (expensive) public-key operations, such as scalar-point multiplications and map-to-point operations, on sensors. We formally prove that our scheme achieves user anonymity as well as authenticated key exchange in an extension of the widely accepted model of Bellare et al. [30]. In proving the security properties, we assume that the cryptographic hash functions used are random oracles and the elliptic curve computational Diffie-Hellman problem is computationally hard. The extended model captures not only the notion of two-factor security but also standard attacks against SCA-WSN schemes like node capture attacks, privileged insider attacks, and stolen verifier attacks.

The remainder of this paper is structured as follows. Section 2 describes an extended security model for the analysis of anonymous SCA-WSN schemes. Section 3 presents our proposed SCA-WSN scheme along with cryptographic primitives on which the security of the scheme relies. Section 4 provides proofs for the security properties of our proposed scheme in the extended security model. Section 5 concludes the paper with a comparative efficiency and security of our scheme and other SCA-WSN schemes.

## A Security Model for Anonymous SCA-WSN Schemes

This section describes a security model extended from the Bellare et al.'s model [30] to analyze authentication and key exchange protocols of anonymous SCA-WSN schemes. Our security model captures the notion of two-factor security as well as the resistance to node capture attacks, privileged insider attacks, stolen verifier attacks, and other common attacks. We provide two security definitions associated with the model, one for authenticated key exchange and one for user anonymity, which collectively define a secure, anonymous SCA-WSN scheme.

### Participants

Let SN and U be the sets of all sensors and users, respectively, registered with the gateway $GW$. Let $E = U \cup SN \cup \{GW\}$. We identify each entity $E \in E$ by a string, and interchangeably use $E$ and $ID_E$ to refer to this identifier string. To formally capture the user anonymity property, we assume that: (1) each user $U \in U$ has its pseudo identity $PID_U$ in addition to the true identity $ID_U$ and (2) the adversary $\mathcal{A}$ is given only $PID_U$ but not $ID_U$.

### Protocol Executions

A user $U \in$ may run multiple sessions of the authentication and key exchange protocol of a SCA-WSN scheme, either serially or concurrently, to establish a session key with a sensor $SN \in$ SN via assistance of the gateway $GW$. Therefore, at any given time, there could be multiple instances of the entities $U$, $SN$ and $GW$. We use $\Pi_E^i$ to denote instance $i$ of entity $E \in E$. Instances of $U$ and $SN$ are said to *accept* when they compute a session key in an execution of the protocol. We denote the session key of $\Pi_E^i$ by $sk_E^i$.

### Long-Lived Keys

During the initialization of the protocol,

- each $U \in U$ chooses its password $PW_U$ from a fixed dictionary D, and

- $GW$ generates its master secret(s), issues a smart card to each $U \in U$, and shares a cryptographic key with each $SN \in SN$.

### Partnering

Informally, two instances are said to be *partners* of each other if they participate together in the same protocol session and as a result, compute the same session key. Formally, partnering between instances is defined in terms of the notion of session identifier. A session identifier (*sid*) is an identifier of a protocol session and is typically defined as a function of the messages exchanged in the session. Let $sid_E^i$ denote the $sid$ of instance $\Pi_E^i$. We say that two instances, $\Pi_U^i$ and $\Pi_{SN}^j$, are partners if (1) both the instances have accepted and (2) $sid_U^i = sid_{SN}^j$.

## Adversary Capabilities

We assume there exists an adversary $\mathcal{A}$ running in a probabilistic polynomial time (PPT) in the security parameter $\kappa$, which represents the bit-length of session keys. We note that the size of the dictionary D is a fixed constant that is independent of the security parameter $\kappa$. The PPT adversary $\mathcal{A}$ has complete control of all communications between entities, can request for access to session keys and long-term keys, and can extract user's information stored on the smart card. These capabilities of $\mathcal{A}$ are modeled via the following oracle queries which are allowed for $\mathcal{A}$ to make.

- Execute($\Pi_U^i$, $\Pi_{SN}^j$, $\Pi_{GW}^k$): This query models passive attacks against the protocol. It prompts an execution of the protocol between the instances $\Pi_U^i$, $\Pi_{SN}^j$ and $\Pi_{GW}^k$, and outputs the transcript of the protocol execution to $\mathcal{A}$.

- Send($\Pi_E^i$, $m$): This query sends a message $m$ to an instance $\Pi_E^i$, modelling active attacks against the protocol. Upon receiving $m$, the instance $\Pi_E^i$ proceeds according to the protocol specification. The message output by $\Pi_E^i$, if any, is returned to $\mathcal{A}$. A query of the form Send($\Pi_U^i$, start:$\langle SN, GW \rangle$) prompts $\Pi_U^i$ to initiate a protocol session with instances of $SN$ and $GW$.

- Reveal($\Pi_E^i$): This query captures the notion of known key security. The instance $\Pi_E^i$, upon receiving the query and if it has accepted, returns the session key, $sk_E^i$, back to $\mathcal{A}$.

- CorruptLL($U$)/CorruptSC($U$): These queries together capture the notion of two-factor security. The former returns the password of $U$ while the latter returns the information stored in the smart card of $U$.

- CorruptLL($SN$): This query returns the long-lived secret(s) of the sensor $SN$, modelling node capture attacks.

- CorruptLL($GW$), modelling privileged insider attacks.

- CorruptVFR($GW$): This query returns the password verifiers stored by $GW$, modelling stolen verifier attacks.

- TestAKE($\Pi_E^i$): This query is used for determining whether the protocol achieves authenticated key exchange or not. If $\Pi_E^i$ has accepted, then depending on a random bit $b$ chosen by the oracle, $\mathcal{A}$ is given either the real session key $sk_E^i$ if $b = 1$ or a random key drawn from the session-key space if $b = 0$.

- TestUA($U$): This query is used for determining whether the protocol provides user anonymity or not. Depending on a randomly chosen bit $b$, $\mathcal{A}$ is given either the identity actually used for $U$ in the protocol sessions (when $b = 1$) or a random identity drawn from the identity space (when $b = 0$).

CorruptLL queries all together also capture the notion of perfect forward secrecy. *SN* and *GW* are said to be corrupted when they are asked a CorruptLL query while *U* is considered as corrupted if it has been asked both CorruptLL and CorruptSC queries.

## Authenticated Key Exchange (AKE)

The AKE security of an authentication and key exchange protocol *P* is defined via the notion of *freshness*. Intuitively, a fresh instance is one that holds a session key which should not be known to the adversary $\mathcal{A}$, and an unfresh instance is one whose session key (or some information about the key) can be known by trivial means. A formal definition of freshness follows:

**Definition 1** (Freshness). *An instance $\Pi_E^i$ is fresh if none of the following occurs*:

1. $\mathcal{A}$ *queries* Reveal($\Pi_E^i$) *or* Reveal($\Pi_{E'}^j$), *where $\Pi_{E'}^j$ is the partner of $\Pi_E^i$.*

2. $\mathcal{A}$ *queries both* CorruptLL($U$) *and* CorruptSC($U$) *when U is E itself or the peer entity of E.*

3. $\mathcal{A}$ *queries* CorruptLL($SN$) *when SN is E itself or the peer entity of E.*

4. $\mathcal{A}$ *queries* CorruptLL($GW$).

Note that this definition of freshness is unable to capture the notion of perfect forward secrecy. (As explained in the next section, the authentication and key exchange protocol of our scheme does not provide perfect forward secrecy.) The AKE security of protocol *P* is defined in the context of the following two-stage experiment:

Experiment **ExpAKE$_0$**:

Stage 1. $\mathcal{A}$ makes any oracle queries at will, except that:

1. $\mathcal{A}$ is not allowed to make the TestAKE($\Pi_E^i$) query if the instance $\Pi_E^i$ is not fresh.

2. $\mathcal{A}$ is not allowed to make the Reveal($\Pi_E^i$) query if it has already made a TestAKE query to $\Pi_E^i$ or its partner instance.

3. $\mathcal{A}$ is not allowed to access to the TestUA oracle.

Stage 2. Once $\mathcal{A}$ decides that Stage 1 is over, it outputs a bit $b'$ as a guess on the hidden bit $b$ chosen by the TestAKE oracle. $\mathcal{A}$ is said to succeed if $b = b'$.

Let SuccAKE$_0$ be the event that $\mathcal{A}$ succeeds in the experiment **ExpAKE$_0$**, and $\mathsf{Adv}_P^{\mathrm{AKE}}(A)$ denote the advantage of $\mathcal{A}$ in breaking the AKE security of protocol *P*. Then, we define $\mathsf{Adv}_P^{\mathrm{AKE}}(A) = 2 \cdot \Pr_{P,A}[\mathsf{SuccAKE}_0] - 1$.

**Definition 2** (AKE Security). *An authentication and key exchange protocol P is AKE-secure if $\mathsf{Adv}_P^{\mathrm{AKE}}(A)$ is negligible for any PPT adversary $\mathcal{A}$.*

## User Anonymity

An authentication and key exchange protocol that does not provide user anonymity may still be rendered AKE-secure. That is, the AKE security does not imply user anonymity. Therefore, a new, separate definition is necessary to capture the user anonymity property. Our definition of user anonymity is based on the notion of cleanness.

**Definition 3** (Cleanness). *A user $U \in \mathbb{U}$ is clean if none of the following occurs*:

1. $\mathcal{A}$ *queries both* CorruptLL($U$) *and* CorruptSC($U$).

2. $\mathcal{A}$ *queries* CorruptLL($GW$).

Note that the definition of cleanness does not impose any restriction on making CorruptLL queries to sensors. This reflects our objective to achieve user anonymity even against sensors.

User anonymity is formalized in the context of the following two-stage experiment: Experiment **ExpUA$_0$**:

Stage 1. $\mathcal{A}$ makes any oracle queries at will, except that:

1. $\mathcal{A}$ is not allowed to make the TestUA($U$) query if the user $U$ is not clean.

2. $\mathcal{A}$ is not allowed to corrupt $GW$ and $U$ if it has already made the TestUA($U$) query.

3. $\mathcal{A}$ is not allowed to access to the TestAKE oracle.

Stage 2. Once $\mathcal{A}$ decides that Stage 1 is over, it outputs a bit $b'$ as a guess on the hidden bit $b$ chosen by the TestUA oracle. $\mathcal{A}$ is said to succeed if $b = b'$.

Let SuccUA$_0$ be the event that $\mathcal{A}$ succeeds in the experiment **ExpUA$_0$**, and $\mathsf{Adv}_P^{\mathrm{UA}}(A)$ denote the advantage of $\mathcal{A}$ in attacking the user anonymity of protocol $P$. Then, we define $\mathsf{Adv}_P^{\mathrm{UA}}(A) = 2 \cdot \Pr_{P,A}[\mathsf{SuccUA}_0] - 1$.

**Definition 4** (User Anonymity). *An authentication and key exchange protocol $P$ provides user anonymity if* $\mathsf{Adv}_P^{\mathrm{UA}}(A)$ *is negligible for any* PPT *adversary $\mathcal{A}$.*

## Our Proposed Scheme

Our SCA-WSN scheme restricts the use of elliptic curve cryptography to anonymous user-to-gateway authentication and thereby allows sensor nodes to perform only lightweight cryptographic operations such as symmetric encryption/decryption, MAC generation/verification, and hash function evaluation. We begin by describing the cryptographic building blocks on which the security of our scheme depends.

### Building Blocks

**Elliptic curve computational Diffie-Hellman (ECCDH) problem.** Let G be an elliptic curve group of prime order $q$. Typically, G will be a subgroup of the group of points on an elliptic curve over a finite field. Any elliptic curve and finite field recommended by NIST [31] can be used to instantiate the group G. The recent work of Choi et al. [20], for example, describes a typical elliptic curve group of a prime order. Let $P$ be a generator of G. The ECCDH problem for G is to compute $xyP \in$ G when given two elements $(xP, yP) \in$ G$^2$, where $x, y \in_R Z_q^*$. We say that the ECCDH assumption holds for G if it is computationally infeasible to solve the ECCDH problem for G. Let $\mathsf{Adv}_G^{\mathrm{ECCDH}}(A)$ be the advantage of an algorithm $\mathcal{A}$ in solving the ECCDH problem for G and be defined as $\mathsf{Adv}_G^{\mathrm{ECCDH}}(A) = \Pr[A(G, P, xP, yP) = xyP]$. We assume that $\mathsf{Adv}_G^{\mathrm{ECCDH}}(A)$ is negligible for all PPT algorithms $\mathcal{A}$ (i.e., the ECCDH assumption holds in G). We denote by $\mathsf{Adv}_G^{\mathrm{ECCDH}}(t)$ the maximum value of $\mathsf{Adv}_G^{\mathrm{ECCDH}}(A)$ over all algorithms $\mathcal{A}$ running in time at most $t$.

**Message authentication code schemes.** A message authentication code (MAC) scheme $\Sigma$ is a pair of efficient algorithms (Mac, Ver) where: (1) the MAC generation algorithm Mac takes as input an $\ell$-bit key $k$ and a message $m$, and outputs a MAC $\sigma$; and (2) the MAC verification algorithm Ver takes as input a key $k$, a message $m$, and a MAC $\sigma$, and outputs 1 if $\sigma$ is valid for message $m$ under the key $k$ or outputs 0 if $\sigma$ is invalid. We require that $\Sigma$ should achieve the strong existential unforgeability against chosen message attacks. To formally define this requirement, let $\mathsf{Adv}_\Sigma^{\mathrm{EF-CMA}}(A)$ be the probability that an adversary $\mathcal{A}$, who mounts an adaptive chosen message attack against $\Sigma$ with oracle access to $\mathsf{Mac}_k(\cdot)$ and $\mathsf{Ver}_k(\cdot)$, outputs a message/

tag pair $(m, \sigma)$ such that: (1) $\mathsf{Ver}_k(m, \sigma) = 1$ and (2) $\sigma$ has not been output by the oracle $\mathsf{Mac}_k(\cdot)$ as a MAC on the message $m$. The, we say that the MAC scheme $\Sigma$ is secure if $\mathsf{Adv}_{\Sigma}^{\mathrm{EF-CMA}}(A)$ is negligible for every PPT adversary $\mathcal{A}$. We use $\mathsf{Adv}_{\Sigma}^{\mathrm{EF-CMA}}(t)$ to denote the maximum value of $\mathsf{Adv}_{\Sigma}^{\mathrm{EF-CMA}}(A)$ over all adversaries $\mathcal{A}$ running in time at most $t$.

**Cryptographic hash functions.**　Let $\kappa$ be the bit-length of session keys, $\ell$ be as defined for $\Sigma$, and $\omega$ be the bit-length of $EID_U$ (see the registration phase of our scheme described in the next section). Then, our scheme uses three cryptographic hash functions $H{:}\{0, 1\}^* \to \{0, 1\}^\kappa$, $J{:}\{0, 1\}^* \to \{0, 1\}^\ell$, and $I{:}\{0, 1\}^* \to \{0, 1\}^\omega$. These hash functions are modelled as random oracles in our security proofs.

**Symmetric encryption schemes.**　A symmetric encryption scheme $\Delta$ is a pair of efficient algorithms $(\mathsf{Enc}, \mathsf{Dec})$ where: (1) the encryption algorithm $\mathsf{Enc}$ takes as input an $\ell$-bit key $k$ and a plaintext message $m$, and outputs a ciphertext $c$; and (2) the decryption algorithm $\mathsf{Dec}$ takes as input a key $k$ and a ciphertext $c$, and outputs a message $m$. For an eavesdropping adversary $\mathcal{A}$ against $\Delta$, and for an integer $n \geq 1$ and a random bit $b \in_R \{0, 1\}$, consider the following indistinguishability experiment where only a single encryption key is used:

Experiment $\mathbf{Exp}_{\Delta}^{\mathrm{IND-SEK}}(A, n, b)$
　　$k \in_R \{0, 1\}^\ell$
　　**for** $i = 1$ **to** $n$
　　　　$(m_{i, 0}, m_{i, 1}) \leftarrow \mathcal{A}(\Delta)$
　　　　$c_i \leftarrow \mathsf{Enc}_k(m_{i, b})$
　　　　$\mathcal{A}(c_i)$
　　$b' \leftarrow \mathcal{A}$, where $b' \in \{0, 1\}$
　　**return** $b'$

We use $\mathsf{Adv}_{\Delta}^{\mathrm{IND-SEK}}(A)$ to denote the advantage of $\mathcal{A}$ in violating the indistinguishability of $\Delta$ in experiment $\mathbf{Exp}_{\Delta}^{\mathrm{IND-SEK}}(A, n, b)$, and define it as

$$\mathsf{Adv}_{\Delta}^{\mathrm{IND-SEK}}(A) = |\Pr[\mathbf{Exp}_{\Delta}^{\mathrm{IND-SEK}}(A, n, 0) = 1] - \Pr[\mathbf{Exp}_{\Delta}^{\mathrm{IND-SEK}}(A, n, 1) = 1]|.$$

We say that the symmetric encryption scheme $\Delta$ is secure if $\mathsf{Adv}_{\Delta}^{\mathrm{IND-SEK}}(A)$ is negligible for every PPT eavesdropper $\mathcal{A}$. Let $\mathsf{Adv}_{\Delta}^{\mathrm{IND-SEK}}(t)$ be the maximum value of $\mathsf{Adv}_{\Delta}^{\mathrm{IND-SEK}}(A)$ over all $\mathcal{A}$ running in time at most $t$.

We now claim that if a symmetric encryption scheme is secure with respect to a single encryption key, then it is also secure with respect to multiple encryption keys. Now consider the following indistinguishability experiment where $d$ encryption keys are used:

Experiment $\mathbf{Exp}_{\Delta}^{\mathrm{IND-SEK}}(A, n, d, b)$
　　**for** $i = 1$ **to** $d$
　　　　$k_i \in_R \{0, 1\}^\ell$
　　　　**for** $j = 1$ **to** $n$
　　　　　　$(m_{i, j, 0}, m_{i, j, 1}) \leftarrow \mathcal{A}(\Delta)$
　　　　　　$c_{i, j} \leftarrow \mathsf{Enc}_{k_i}(m_{i, j, b})$
　　　　　　$\mathcal{A}(c_{i, j})$
　　　　$b' \leftarrow \mathcal{A}$, where $b' \in \{0, 1\}$
　　　　**return** $b'$

We define $\mathsf{Adv}_{\Delta}^{\mathrm{IND-MEK}}(A)$ and $\mathsf{Adv}_{\Delta}^{\mathrm{IND-MEK}}(t)$ respectively as

$$\mathsf{Adv}_{\Delta}^{\mathrm{IND-MEK}}(A) = |\Pr[\mathbf{Exp}_{\Delta}^{\mathrm{IND-MEK}}(A, n, d, 0) = 1] - \Pr[\mathbf{Exp}_{\Delta}^{\mathrm{IND-MEK}}(A, n, d, 1) = 1]|$$

and

$$\mathsf{Adv}_{\Delta}^{\mathrm{IND-MEK}}(t) = \max_{A} \left\{ \mathsf{Adv}_{\Delta}^{\mathrm{IND-MEK}}(A) \right\},$$

where the maximum is over all $\mathcal{A}$ running in time at most $t$.

**Lemma 1**. *For any symmetric encryption scheme $\Delta$,*

$$\mathsf{Adv}_{\Delta}^{\mathrm{IND-MEK}}(t) \leq d \cdot \mathsf{Adv}_{\Delta}^{\mathrm{IND-SEK}}(t),$$

*where $d$ is as defined for experiment* $\mathbf{Exp}_{\Delta}^{\mathrm{IND-MEK}}(A, n, d, b)$.

*Proof.* Assume an adversary $\mathcal{A}$ who attacks the indistinguishability of $\Delta$ in $\mathbf{Exp}_{\Delta}^{\mathrm{IND-MEK}}(A, n, d, b)$ with time complexity $t$. The proof proceeds with a standard hybrid argument [32]. Consider a sequence of $d + 1$ hybrid experiments $\mathbf{Exp}_{\Delta, \xi}^{\mathrm{IND-MEK}}(A, n, d, b)$, $0 \leq \xi \leq d$, where each $\mathbf{Exp}_{\Delta, \xi}^{\mathrm{IND-MEK}}(A, n, d, b)$ is different from $\mathbf{Exp}_{\Delta}^{\mathrm{IND-MEK}}(A, n, d, b)$ only in that each $c_{i,j}$ is set as follows:

$$c_{i,j} \leftarrow \begin{cases} \mathsf{Enc}_{k_i}(m_{i,j,1}) & \text{if } i \leq \xi \\ \mathsf{Enc}_{k_i}(m_{i,j,0}) & \text{otherwise.} \end{cases}$$

The experiments $\mathbf{Exp}_{\Delta, 0}^{\mathrm{IND-MEK}}(A, n, d, b)$ and $\mathbf{Exp}_{\Delta, d}^{\mathrm{IND-MEK}}(A, n, d, b)$ at the extremes of the sequence are identical to the experiments $\mathbf{Exp}_{\Delta}^{\mathrm{IND-MEK}}(A, n, d, 0)$ and $\mathbf{Exp}_{\Delta}^{\mathrm{IND-MEK}}(A, n, d, 1)$, respectively. As we move from $\mathbf{Exp}_{\Delta, \xi-1}^{\mathrm{IND-MEK}}(A, n, d, b)$ to $\mathbf{Exp}_{\Delta, \xi}^{\mathrm{IND-MEK}}(A, n, d, b)$ in the sequence, we change the $n$ ciphertexts $c_{\xi,1}, \ldots, c_{\xi,n}$ from encryptions of the first plaintexts to encryptions of the second plaintexts. Since there are $d$ such moves from $\mathbf{Exp}_{\Delta, 0}^{\mathrm{IND-MEK}}(A, n, d, b)$ to $\mathbf{Exp}_{\Delta, d}^{\mathrm{IND-MEK}}(A, n, d, b)$, the inequality of the lemma follows immediately if we prove that the difference between the probabilities that $\mathcal{A}$ outputs 1 in any two neighboring experiments $\mathbf{Exp}_{\Delta, \xi-1}^{\mathrm{IND-MEK}}(A, n, d, b)$ and $\mathbf{Exp}_{\Delta, \xi}^{\mathrm{IND-MEK}}(A, n, d, b)$ is at most $\mathsf{Adv}_{\Delta}^{\mathrm{IND-SEK}}(t)$. That is, to complete the proof, it suffices to show that for any $1 \leq \xi \leq d$,

$$|\Pr[\mathbf{Exp}_{\Delta, \xi-1}^{\mathrm{IND-MEK}}(A, n, d, b) = 1] - \Pr[\mathbf{Exp}_{\Delta, \xi}^{\mathrm{IND-MEK}}(A, n, d, b) = 1]| \leq \mathsf{Adv}_{\Delta}^{\mathrm{IND-SEK}}(t). \quad (1)$$

Let $\epsilon = | \Pr[\mathbf{Exp}_{\Delta, \xi-1}^{\mathrm{IND-MEK}}(A, n, d, b) = 1] - \Pr[\mathbf{Exp}_{\Delta, \xi}^{\mathrm{IND-MEK}}(A, n, d, b) = 1] |$. Then, to prove Equation 1, we will construct, from $\mathcal{A}$, an adversary $\mathcal{A}_{\xi}$ who attacks the indistinguishability of $\Delta$ in $\mathbf{Exp}_{\Delta}^{\mathrm{IND-SEK}}(A, n, b)$ with advantage $\epsilon$.

$\mathcal{A}_{\xi}$ begins by invoking adversary $\mathcal{A}$, then proceeds to simulate the indistinguishability experiment for $\mathcal{A}$, and finally ends by outputting whatever bit $\mathcal{A}$ eventually outputs. In the simulated experiment, $\mathcal{A}_{\xi}$ generates the ciphertexts exactly as in the hybrid experiment $\mathbf{Exp}_{\Delta, \xi}^{\mathrm{IND-MEK}}(A, b, n)$ except that it generates $c_{\xi,1}, \ldots, c_{\xi,n}$ as follows:

When $\mathcal{A}$ outputs the $n$ plaintext pairs $(m_{\xi,1,0}, m_{\xi,1,1}), \ldots, (m_{\xi,n,0}, m_{\xi,n,1})$, $\mathcal{A}_{\xi}$ outputs them as its own plaintext pairs in experiment $\mathbf{Exp}_{\Delta}^{\mathrm{IND-SEK}}(A_{\xi}, n, b)$, receives in return the ciphertexts $c_1, \ldots, c_n$, and sets $c_{\xi,1} = c_1, \ldots, c_{\xi,n} = c_n$.

Then, it follows that:

- the probability that $\mathcal{A}_{\xi}$ outputs 1 when the given ciphertexts are the encryptions of the first plaintexts is equal to the probability that $\mathcal{A}$ outputs 1 in the experiment $\mathbf{Exp}_{\Delta, \xi-1}^{\mathrm{IND-MEK}}(A, n, d, b)$, and

- the probability that $\mathcal{A}_\xi$ outputs 1 when the given ciphertexts are the encryptions of the second plaintexts is equal to the probability that $\mathcal{A}$ outputs 1 in the experiment $\mathbf{Exp}_{\Delta,\xi}^{\text{IND-MEK}}(A, n, d, b)$.

  That is:

  $$\mathsf{Adv}_\Delta^{\text{IND-SEK}}(A_\xi) = |\Pr[\mathbf{Exp}_{\Delta,\xi-1}^{\text{IND-MEK}}(A, n, d, b) = 1] - \Pr[\mathbf{Exp}_{\Delta,\xi}^{\text{IND-MEK}}(A, n, d, b) = 1]|.$$

Since $\mathcal{A}_\xi$ has time complexity $t$, it follows that $\mathsf{Adv}_\Delta^{\text{IND-SEK}}(A_\xi) \leq \mathsf{Adv}_\Delta^{\text{IND-SEK}}(t)$ by definition. This completes the proof of Equation 1 and hence the proof of Lemma 1.

## Description of the Scheme

The scheme consists of three phases: the registration phase, the authentication and key exchange phase, and the password update phase. During the system initialization, the gateway $GW$ determines the following public parameters: (1) an elliptic curve group G with a generator $P$ of prime order $q$, (2) a MAC scheme $\Sigma = (\mathsf{Mac}, \mathsf{Ver})$, (3) a symmetric encryption scheme $\Delta = (\mathsf{Enc}, \mathsf{Dec})$, and (4) three hash functions $H$, $J$ and $I$. We assume that these parameters are known to all parties in the network including the adversary $\mathcal{A}$. As part of the system initialization, $GW$ chooses two master secrets $y \in Z_q^*$ and $z \in \{0, 1\}^\ell$, computes its public key $Y = yP$, and shares a secret key $k_{GS} = J(ID_{SN}\|z)$ with each sensor $SN$.

**Registration phase.** A user $U$ should register itself with the gateway $GW$ before it can ever gain access to the sensor network and data. The registration proceeds as follows:

1. $U$ chooses its identity $ID_U$ and password $PW_U$ at will, and submits the identity $ID_U$ to $GW$ via a secure channel.

2. $GW$ computes $EID_U = \mathsf{Enc}_z(ID_U\|ID_{GW})$ and issues $U$ a smart card loaded with $\{EID_U, Y, ID_{GW}, G, P, \Sigma, \Delta, H, J, I\}$. (We assume that $q$ is implicit in G.)

3. $U$ replaces $EID_U$ with $XEID_U = EID_U \oplus I(ID_U\|PW_U)$.

**Authentication and key exchange phase.** $U$ needs to perform this phase with $SN$ and $GW$ whenever it wishes to access to the sensor network and data. The steps of the phase are depicted in Fig. 1 and are described as follows:

**Step 1**. $U$ inserts its smart card into a card reader and inputs its identity $ID_U$ and password $PW_U$. Then, the smart card retrieves the current timestamp $T_U$, selects two random $x \in Z_q^*$ and $k_{US} \in \{0, 1\}^\kappa$, and computes

$$
\begin{aligned}
X &= xP, \\
K_{UG} &= xY, \\
k_{UG} &= J(T_U \parallel X \parallel Y \parallel K_{UG}), \\
EID_U &= XEID_U \oplus I(ID_U \parallel PW_U), \\
C_U &= \mathsf{Enc}_{k_{UG}}(ID_U \parallel EID_U \parallel k_{US}), \\
\sigma_U &= \mathsf{Mac}_{k_{UG}}(ID_{GW} \parallel ID_{SN} \parallel T_U \parallel C_U).
\end{aligned}
$$

After the computations, the smart card sends the message $M_1 = \langle T_U, ID_{SN}, X, C_U, \sigma_U \rangle$ to the gateway $GW$.

**Step 2**. $GW$ rejects the message $M_1$ (and aborts the session) if $T_U$ is not fresh. Otherwise, $GW$ computes $K_{UG} = yX$ and $k_{UG} = J(T_U\|X\|Y\|K_{UG})$, and checks if $\mathsf{Ver}_{k_{UG}}(ID_{GW}\|ID_{SN}\|$

$\boxed{U}$                                        $\boxed{GW}$                                        $\boxed{SN}$

inputs $ID_U$ and $PW_U$
retrieves the timestamp $T_U$
$x \in_R \mathbb{Z}_q^*,\ k_{US} \in \{0,1\}^\kappa$
$X = xP$
$K_{UG} = xY$
$k_{UG} = J(T_U \| X \| Y \| K_{UG})$
$EID_U = XEID_U \oplus I(ID_U \| PW_U)$
$C_U = \mathsf{Enc}_{k_{UG}}(ID_U \| EID_U \| k_{US})$
$\sigma_U = \mathsf{Mac}_{k_{UG}}(ID_{GW} \| ID_{SN} \| T_U \| C_U)$

$\xrightarrow{\hspace{1cm} M_1 = \langle T_U, ID_{SN}, X, C_U, \sigma_U \rangle \hspace{1cm}}$

checks the freshness of $T_U$
$K_{UG} = yX$
$k_{UG} = J(T_U \| X \| Y \| K_{UG})$
$\mathsf{Ver}_{k_{UG}}(ID_{GW} \| ID_{SN} \| T_U \| C_U, \sigma_U) \stackrel{?}{=} 1$
$\mathsf{Dec}_{k_{UG}}(C_U)$ and $\mathsf{Dec}_z(EID_U)$ yield the same $ID_U$ ?
retrieves the timestamp $T_{GW}$
$C_{GW} = \mathsf{Enc}_{k_{GS}}(k_{US})$
$\sigma_{GW} = \mathsf{Mac}_{k_{GS}}(ID_{GW} \| ID_{SN} \| T_{GW} \| T_U \| C_{GW})$

$\xrightarrow{\hspace{1cm} M_2 = \langle ID_{GW}, T_{GW}, T_U, C_{GW}, \sigma_{GW} \rangle \hspace{1cm}}$

checks the freshness of $T_{GW}$
$\mathsf{Ver}_{k_{GS}}(ID_{GW} \| ID_{SN} \| T_{GW} \| T_U \| C_{GW}, \sigma_{GW}) \stackrel{?}{=} 1$
$k_{US} = \mathsf{Dec}_{k_{GS}}(C_{GW})$
$sk = H(k_{US} \| T_U \| ID_{SN})$
$\rho_{SN} = H(k_{US} \| ID_{SN} \| T_U)$

$\xleftarrow{\hspace{1cm} M_3 = \langle \rho_{SN} \rangle \hspace{1cm}}$

$\rho_{SN} \stackrel{?}{=} H(k_{US} \| ID_{SN} \| T_U)$
$sk = H(k_{US} \| T_U \| ID_{SN})$

**Fig 1. Our proposed authentication and key exchange protocol for wireless sensor networks.**

$T_U \| C_U, \sigma_U) = 1$. If the check fails, $GW$ aborts the session. Otherwise, $GW$ decrypts $C_U$ with key $k_{UG}$ and then $EID_U$ with key $z$, and checks if the decryption of $EID_U$ yields the same $ID_U$ as produced through the decryption of $C_U$. If only the two IDs match, $GW$ retrieves the current timestamp $T_{GW}$, computes

$$
\begin{aligned}
C_{GW} &= \mathsf{Enc}_{k_{GS}}(k_{US}), \\
\sigma_{GW} &= \mathsf{Mac}_{k_{GS}}(ID_{GW} \| ID_{SN} \| T_{GW} \| T_U \| C_{GW}),
\end{aligned}
$$

and sends the message $M_2 = \langle ID_{GW}, T_{GW}, T_U, C_{GW}, \sigma_{GW} \rangle$ to the sensor $SN$.

**Step 3.** Upon receiving $M_2$, $SN$ verifies that (1) $T_{GW}$ is fresh and (2) $\mathsf{Ver}_{k_{GS}}(ID_{GW} \| ID_{SN} \| T_{GW} \| T_U \| C_{GW}, \sigma_{GW}) = 1$. If any of the verifications fails, $SN$ aborts the session. Otherwise, $SN$ decrypts $C_{GW}$ to obtain $k_{US}$ and computes the session key $sk$ and the authenticator $\rho_{SN}$ as follows:

$$
\begin{aligned}
sk &= H(k_{US} \| T_U \| ID_{SN}), \\
\rho_{SN} &= H(k_{US} \| ID_{SN} \| T_U).
\end{aligned}
$$

Then, $SN$ sends the message $M_3 = \langle \rho_{SN} \rangle$ to the user $U$.

**Step 4.** With $M_3$ in hand, $U$ checks if $\rho_{SN}$ is equal to $H(k_{US} \| ID_{SN} \| T_U)$. $U$ aborts the session if the check fails or otherwise computes the session key $sk = H(k_{US} \| T_U \| ID_{SN})$.

**Password update phase.** One of the recommended guidelines for achieving better password security is to enforce regular password updates. In our scheme, users can change their passwords either non-interactively or interactively. The non-interactive password change procedure proceeds as follows:

1. $U$ inserts his smart card into a card reader and enters the identity $ID_U$, the current password $PW_U$, and the new password $PW'_U$.

2. The smart card computes $XEID'_U = XEID_U \oplus I(ID_U \| PW_U) \oplus I(ID_U \| PW'_U)$ and replaces $XEID_U$ with $XEID'_U$.

Although this procedure is simple and non-interactive, it may render the smart card unusable if the user enters a wrong password by mistake or an adversary intentionally inputs an arbitrary password after gaining temporary access to the smart card. When an invalid password is entered, subsequent login requests of the user will be rejected unless it reregisters with the gateway. This problem can be addressed by storing a password verifier on the smart card, which is used to check the correctness of the user-given password. However, as soon as the smart card contains a password verifier, the scheme becomes vulnerable to an offline dictionary attack under the non-tamper-resistance assumption of smart cards and, consequently, fails to achieve two-factor security. This is clearly unacceptable and, therefore, we suggest the following interactive password change procedure.

1. $U$ inserts his smart card into a card reader and enters the identity $ID_U$, the current password $PW_U$, and the new password $PW'_U$.

2. The smart card retrieves the current timestamp $T_U$, selects a random $x \in Z_q^*$, and computes

$$
\begin{aligned}
X &= xP, \\
K_{UG} &= xY, \\
k_{UG} &= J(T_U \parallel X \parallel Y \parallel K_{UG}), \\
EID_U &= XEID_U \oplus I(ID_U \parallel PW_U), \\
C_U &= \mathsf{Enc}_{k_{UG}}(ID_U \parallel EID_U).
\end{aligned}
$$

The smart card sends a password update request $\langle T_U, X, C_U \rangle$ to the gateway $GW$.

3. $GW$ rejects the request if $T_U$ is not fresh. Otherwise, $GW$ computes $K_{UG} = yX$ and $k_{UG} = J$ $(T_U \parallel X \parallel Y \parallel K_{UG})$, decrypts $C_U$ with key $k_{UG}$ and then $EID_U$ with key $z$, and checks whether the two decryptions return the same $ID_U$. If the check succeeds, $GW$ computes $\rho_{GW} = H$ $(ID_{GW} \parallel ID_U \parallel X \parallel k_{UG})$ and sends it to the smart card. Otherwise, $GW$ sends a failure message to the smart card.

4. The smart card aborts the password change procedure if it receives a failure message or $\rho_{GW}$ is not equal to $H(ID_{GW} \parallel ID_U \parallel X \parallel k_{UG})$. Otherwise, it sets $XEID_U = EID_U \oplus I(ID_U \parallel PW_U')$.

This interactive password change procedure provides a secure yet practical way of updating user password, though it is more expensive than the non-interactive one.

## Performance and Security Comparison

In Table 1, we provide a comparative summary between our scheme and other SCA-WSN schemes both in terms of computation and security. As shown in the table, our scheme requires the sensor $SN$ to perform only lightweight cryptographic operations while enjoying provable anonymity in an extension of the widely accepted model of Bellare et al. [30]. While the recent schemes of Shi & Gong [12] and Choi et al. [20] provide forward secrecy, they impose 2 scalar-

**Table 1. A comparative summary of smart-card-based user authentication schemes for wireless sensor networks.**

| Scheme | Computation | | Security | | | | |
|---|---|---|---|---|---|---|---|
| | *SN* | *U+SN+GW* | **SKS** | **UA** | **FS** | **RSSC** | **RNC** |
| Jiang et al. [19] | 5H | 22H | Yes | No | No | No | Yes |
| Khan & Kumari [18] | 7H | 3E+20H | Yes | No | No | No | Yes |
| Kim et al. [17] | 2H | 18H | Yes | No | No | No | Yes |
| Chi et al. [16] | 2E+1A+1H | 4E+1A+5H | Yes | No | No | No | Yes |
| He et al. [15] | 2E+1H | 10E+7H | Yes | No | No | No | Yes |
| Kumar et al. [14] | 2E+1H | 7E+8H | Yes | No | No | No | Yes |
| Li et al. [13] | 6H | 26H | Yes | No | No | No | Yes |
| Xue et al. [11] | 6H | 22H | Yes | No | No | No | Yes |
| Vaidya et al. [10] | 2H | 15H | Yes | No | No | No | No [17] |
| **Our scheme** | 1E+1A+2H | 3M+5E+4A+7H | Yes | Proven | No | Yes | Yes |
| Choi et al. [20] | 2M+5H | 6M+18H | Yes | No | Yes | No | Yes |
| Shi & Gong [12] | 2M+4H | 6M+15H | No [20] | No | Yes | No | Yes |
| Yeh et al. [6] | 2M+1P+2H | 8M+2P+9H | No [33] | No | No [33] | No | Yes |

SKS: session key security; UA: user anonymity; FS: forward secrecy; RSSC: resistance to stolen smart card attacks; RNC: resistance to node capture attacks. $M$: scalar-point multiplication; $P$: map-to-point operation; $E$: symmetric encryption/decryption; $A$: MAC generation/verification; $H$: hash function evaluation.

doi:10.1371/journal.pone.0116709.t001

**Table 2. Crypto++ 5.6.0 benchmarks for SHA-1, HMAC and AES.**

| Algorithm | SHA-1 | HMAC (SHA-1) | AES (with 128-bit key) | | | | |
|---|---|---|---|---|---|---|---|
| | | | CTR | CBC | OFB | CFB | ECB |
| Cycles Per Byte | 11.4 | 11.9 | 12.6 | 16.0 | 16.9 | 16.1 | 16.0 |

point multiplications on the resource-constrained sensor *SN*. Note that scalar-point multiplication is much more expensive than the lightweight cryptographic operations considered in the table, such as symmetric encryption/decryption, MAC generation/verification, and hash function evaluation. Moreover, these two schemes fail to achieve user anonymity despite their use of elliptic curve cryptography. The schemes presented in [10, 11, 13–19] are computationally efficient, but suffer from the inherent failure of user anonymity. To the best of our knowledge, all existing SCA-WSN schemes fall into one of the two classes.

According to Crypto++ 5.6.0 benchmarks that ran on an Intel Core 2 1.83 GHz processor under Windows Vista in 32-bit mode, SHA-1 and HMAC take 11.4 and 11.9 cycles per byte respectively; while AES (with 128-bit key) takes 12.6 to 16.9 cycles per byte, depending on the operation mode used—see Table 2 and we refer interested readers to http://www.cryptopp.com/benchmarks.html for Crypto++ benchmarks for commonly used cryptographic algorithms.

Our scheme requires the sensor *SN* to perform $1E+1A+2H$ operations which amount to about $4.5H$ operations. Therefore, in terms of computational requirements for *SN*, our scheme is comparable with other SCA-WSN schemes [11, 13–16, 18, 19] using only lightweight cryptographic techniques. Although the schemes of Vaidya et al. [10] and Kim et al. [17] require *SN* to perform only 2 hash function evaluations, these schemes do not achieve user anonymity and are vulnerable to a stolen smart card attack. Under the non-tamper-resistance assumption of smart cards, our scheme is the only one that provides user anonymity and resists stolen smart card attacks.

## Security Proofs

We now prove that the authentication and key exchange protocol of our scheme is AKE-secure (in the sense of Definition 2) and provides user anonymity (in the sense of Definition 4). Recall that the security model described in Section 2 captures various SCA-WSN specific attacks (such as stolen smart card attacks, node capture attacks, privileged insider attacks, and stolen verifier attacks) as well as other common attacks (like impersonation attacks, man-in-the-middle attacks, replay attacks, and known key attacks) [21, 23, 25, 34]. Before providing formal security proofs in the model, we briefly discuss the security of our scheme against SCA-WSN specific attacks.

**Stolen smart card attacks**. Our scheme does not require a password verifier to be stored on the smart card of user *U*. Moreover, even if an adversary managed to obtain the ciphertext $C_U = \mathsf{Enc}_{k_{UG}}(ID_U\|EID_U\|k_{US})$, the adversary would be unable to exploit $C_U$ as a password verifier since, under the ECCDH assumption, it is infeasible to compute $k_{UG} = J(T_U\|X\|Y\|K_{UG})$ from $X$ and $Y$. Thus, our scheme is resistant against stolen smart card attacks.

**Node capture attacks**. In our scheme, each sensor node *SN* holds its individual secret key $k_{GS} = J(ID_{SN}\|z)$ which is shared only with the gateway *GW*. In other words, different sensor nodes have different secret keys (with an overwhelming probability). Thus, the secret key $k_{GS}$ obtained by capturing a sensor node *SN* will be of no use in impersonating another sensor node $SN'$ who holds a secret key other than $k_{GS}$. Therefore, node capture attacks are not possible against our scheme.

**Privileged insider attacks.** A privileged insider attack occurs when the gateway administrator can access a user's password to impersonate the user. In our scheme, the gateway $GW$ receives no password-related information from the user $U$ and does not manage any table for storing such information. It is thus clear that privileged insider attacks cannot be mounted against our scheme.

**Stolen verifier attacks.** In a stolen verifier attack, the adversary attempts to impersonate a legitimate user by stealing the user's password verifier stored on the gateway $GW$. However, in our scheme, $GW$ does not store a password verifier of any kind but stores only two master secrets $y$ and $z$ which are selected independently of user passwords. Hence, our scheme is secure against stolen verifier attacks.

## User Anonymity

**Theorem 1.** *Our authentication and key exchange protocol, P, provides user anonymity in the random oracle model under the ECCDH assumption in* G *and the security of the symmetric encryption scheme* $\Delta$.

*Proof.* Let $\mathcal{A}$ be a PPT adversary against the user anonymity property of protocol $P$. We prove the theorem by making a series of modifications to the original experiment **ExpUA$_0$**, bounding the difference in the success probability of $\mathcal{A}$ between two consecutive experiments, and ending up with an experiment where $\mathcal{A}$ has a success probability of 1/2 (i.e., $\mathcal{A}$ has no advantage). Let SuccUA$_i$ denote the event that $\mathcal{A}$ correctly guesses the hidden bit $b$ chosen by the TestUA oracle in experiment **ExpUA$_i$**. Let $t'_{UA}$ be the maximum time required to perform the experiment **ExpUA$_i$** involving the adversary $\mathcal{A}$.

**Experiment ExpUA$_1$.** In this experiment, we simulate the random oracle $J$ as follows:

Simulation of the $J$ oracle: For each $J$ query on a string $str$, the simulator first checks if an entry of the form $(str, j)$ is in a list called JList which contains all the input-output pairs of $J$. If such an entry exists in JList, the simulator returns $j$ as the output of the $J$ query. Otherwise, the simulator chooses a random $\ell$-bit string $j'$, returns $j'$ in response to the query, and adds the entry $(str, j')$ to JList.

For all other oracle queries of $\mathcal{A}$, the simulator answers them as in the original experiment **ExpUA$_0$**. Then, **ExpUA$_1$** is perfectly indistinguishable from **ExpUA$_0$** and therefore, Claim 1 holds.

**Claim 1.** $\Pr_{P,A}[\text{SuccUA}_1] = \Pr_{P,A}[\text{SuccUA}_0]$.

**Experiment ExpUA$_2$.** Here, we modify the experiment so that $X$ is computed as follows: The **ExpUA$_2$** modification:

- The simulator chooses a random exponent $a \in Z_q^*$ and computes $A = aP$.

- For each user instance, the simulator chooses a random $r \in Z_q^*$ and sets $X = rA$.

As a result of the modification, each $K_{UG}$ is set to $rayP$ for some random $r \in Z_q^*$. Since the view of $\mathcal{A}$ is identical between **ExpUA$_2$** and **ExpUA$_1$**, it follows that:

**Claim 2.** $\Pr_{P,A}[\text{SuccUA}_2] = \Pr_{P,A}[\text{SuccUA}_1]$.

**Experiment ExpUA$_3$.** We next modify the computations of $X$ and $Y$ as follows: The **ExpUA$_3$** modification:

- The simulator chooses two random elements $A, B \in$ G and sets $Y = B$.

- For each instance of clean users, the simulator chooses a random $r \in Z_q^*$ and sets $X = rA$. For other instances, the simulator computes $X$ as in experiment **ExpUA$_2$**.

- For each instance of clean users, the simulator sets each $k_{UG}$ to a random $\ell$-bit string. For other instances, the simulator computes $k_{UG}$ as in experiment **ExpUA$_2$**.

Since $k_{UG}$ is set to a random $\ell$-bit string (for instances of clean users), the success probability of $\mathcal{A}$ may be different between **ExpUA$_3$** and **ExpUA$_2$** if it makes an $J(T_U\|X\|Y\|K_{UG})$ query. However, this difference is bounded by Claim 3.

**Claim 3.** $| \operatorname{Pr}_{P,\mathcal{A}}[\mathsf{SuccUA}_3] - \operatorname{Pr}_{P,\mathcal{A}}[\mathsf{SuccUA}_2] | \leq 1/q_J \cdot \mathsf{Adv}_G^{\mathrm{ECCDH}}(t_{UA}^3)$, where $q_J$ is the number of queries made to the $J$ oracle.

*Proof.* We prove the claim via a reduction from the ECCDH problem which is believed to be hard. Assume that the success probability of $\mathcal{A}$ is non-negligibly different between **ExpUA$_3$** and **ExpUA$_2$**. Then we construct an algorithm $\mathcal{A}_{\mathrm{ECCDH}}$ that solves the ECCDH problem in G with a non-negligible advantage. The objective of $\mathcal{A}_{\mathrm{ECCDH}}$ is to compute and output the value $W = uvP \in \mathrm{G}$ when given an ECCDH-problem instance $(U = uP, V = vP) \in \mathrm{G}^2$. $\mathcal{A}_{\mathrm{ECCDH}}$ runs $\mathcal{A}$ as a subroutine while simulating all the oracles on its own.

$\mathcal{A}_{\mathrm{ECCDH}}$ handles all the oracle queries of $\mathcal{A}$ as specified in experiment **ExpUA$_3$** but using $U$ and $V$ in place of $X$ and $Y$. When $\mathcal{A}$ outputs its guess $b'$, $\mathcal{A}_{\mathrm{ECCDH}}$ chooses an entry of the form $(T_U\|X\|Y\|K,j)$ at random from JList and terminates outputting $K/r$. From the simulation, it is clear that $\mathcal{A}_{\mathrm{ECCDH}}$ outputs the desired result $W = uvP$ with probability at least $1/q_J$ if $\mathcal{A}$ makes a $J(T_U\|X\|Y\|K_{UG})$ query for some instance of a clean user $U \in \mathrm{U}$. This completes the proof of Claim 3.

**Experiment ExpUA$_4$.** We finally modify the experiment so that, for each clean user $U \in \mathrm{U}$, a random identity $ID_U'$ drawn from the identity space is used in place of the true identity $ID_U$ in generating $C_U$.

**Claim 4.** $| \operatorname{Pr}_{P,\mathcal{A}}[\mathsf{SuccUA}_4] - \operatorname{Pr}_{P,\mathcal{A}}[\mathsf{SuccUA}_3] | \leq \mathsf{Adv}_\Delta^{\mathrm{IND-MEK}}(t_{UA}^4)$.

*Proof.* We prove the claim by constructing an eavesdropping adversary $\mathcal{A}_{\mathrm{IND-MEK}}$ who attacks the indistinguishability of $\Delta$ in $\mathbf{Exp}_\Delta^{\mathrm{IND-MEK}}(A, n, d, b)$ with advantage equal to $| \operatorname{Pr}_{P,\mathcal{A}}[\mathsf{SuccUA}_4] - \operatorname{Pr}_{P,\mathcal{A}}[\mathsf{SuccUA}_3] |$ (see Section 1 for details of experiment $\mathbf{Exp}_\Delta^{\mathrm{IND-MEK}}(A, n, d, b)$).

$\mathcal{A}_{\mathrm{IND-MEK}}$ begins by choosing a random bit $b \in \{0, 1\}$. Then, $\mathcal{A}_{\mathrm{IND-MEK}}$ invokes the adversary $\mathcal{A}$ and answers all the oracle queries of $\mathcal{A}$ as in experiment **ExpUA$_3$** except that, for each clean user $U \in \mathrm{U}$, it generates $C_U$ by accessing its own encryption oracle as follows:

$\mathcal{A}_{\mathrm{IND-MEK}}$ outputs $(ID_U\|EID_U\|k_{US}, ID_U'\|EID_U\|k_{US})$ as the first plaintext-pair in the indistinguishability experiment $\mathbf{Exp}_\Delta^{\mathrm{IND-MEK}}$. Let $c_1$ be the ciphertext received in return for the first pair. $\mathcal{A}_{\mathrm{IND-MEK}}$ sets $C_U$ equal to the ciphertext $c_1$.

That is, $\mathcal{A}_{\mathrm{IND-MEK}}$ sets $C_U$ to the encryption of either $ID_U\|EID_U\|k_{US}$ or $ID_U'\|EID_U\|k_{US}$. Now when $\mathcal{A}$ terminates and outputs its guess $b'$, $\mathcal{A}_{\mathrm{IND-MEK}}$ outputs 1 if $b = b'$, and 0 otherwise. Then, it is clear that:

- the probability that $\mathcal{A}_{\mathrm{IND-MEK}}$ outputs 1 when the first plaintexts are encrypted in the experiment $\mathbf{Exp}_\Delta^{\mathrm{IND-MEK}}$ is equal to the probability that $\mathcal{A}$ succeeds in the experiment **ExpUA$_3$**, and

- the probability that $\mathcal{A}_{\mathrm{IND-MEK}}$ outputs 1 when the second plaintexts are encrypted in the experiment $\mathbf{Exp}_\Delta^{\mathrm{IND-MEK}}$ is equal to the probability that $\mathcal{A}$ succeeds in the experiment **ExpUA$_4$**.

That is, $\mathsf{Adv}_{\Delta}^{\text{IND−MEK}}(A_{\text{IND−MEK}}) = | \Pr_{P,A}[\mathsf{SuccUA_4}] - \Pr_{P,A}[\mathsf{SuccUA_3}] |$. Note that in the simulation, $\mathcal{A}_{\text{IND-MEK}}$ eavesdrops at most $q_{\text{send}}$ encryptions which is polynomial in the security parameter $\ell$. This completes the proof of Claim 4.

In the experiment **ExpUA₄**, the adversary $\mathcal{A}$ gains no information on the hidden bit $b$ chosen by the TestUA oracle because the identities of all clean users are chosen uniformly at random from the identity space. It, therefore, follows that $\Pr_{P,A}[\mathsf{SuccUA_4}] = 1/2$. This result combined with Claims 1–4 yields the statement of Theorem 1.

## AKE Security

**Theorem 2**. *As long as the MAC scheme $\Sigma$ and the symmetric encryption scheme $\Delta$ are both secure, our authentication and key exchange protocol P is secure in the random oracle model under the ECCDH assumption in* G.

*Proof.* Fix a PPT adversary $\mathcal{A}$ against the security of the protocol $P$. To prove the theorem, we make a series of modifications to the original experiment **ExpAKE₀**, bounding the effect of each change in the experiment on the success probability of $\mathcal{A}$ and ending up with an experiment where $\mathcal{A}$ has a success probability of 1/2. We use $\mathsf{SuccAKE_i}$ to denote the event that $\mathcal{A}$ correctly guesses the hidden bit $b$ chosen by the Test oracle in experiment **ExpAKE_i**. Let $t^i_{AKE}$ be the maximum time required to perform the experiment **ExpAKE_i** involving the adversary $\mathcal{A}$.

**Experiment ExpAKE₁**. This experiment is different from **ExpAKE₀** in that the random oracle $J$ is simulated as follows:

Simulation of the $J$ oracle: For each $J$ query on a string $str$, the simulator first checks if an entry of the form $(str,j)$ is in a list called JList which contains all the input-output pairs of $J$. If such an entry exists in JList, the simulator returns $j$ as the output of the $J$ query. Otherwise, the simulator chooses a random $\ell$-bit string $j'$, returns $j'$ in response to the query, and adds the entry $(str,j')$ to JList.

The other oracle queries of $\mathcal{A}$ are answered as in the original experiment **ExpAKE₀**. Then, since $J$ is a random oracle, **ExpAKE₁** is perfectly indistinguishable from **ExpAKE₀**, and Claim 5 immediately follows.

**Claim 5**. $\Pr_{P,A}[\mathsf{SuccAKE_1}] = \Pr_{P,A}[\mathsf{SuccAKE_0}]$.

**Experiment ExpAKE₂**. Here, we modify the experiment so that $X$ is computed as follows: The **ExpAKE₂** modification:

- The simulator chooses a random exponent $a \in Z_q^*$ and computes $A = aP$.

- For each instance of users, the simulator chooses a random $r \in Z_q^*$ and sets $X = rA$.

As a result, each $K_{UG}$ is set to $rayP$ for some random $r \in Z_q^*$. Since the view of $\mathcal{A}$ is identical between **ExpAKE₂** and **ExpAKE₁**, it follows that:

**Claim 6**. $\Pr_{P,A}[\mathsf{SuccAKE_2}] = \Pr_{P,A}[\mathsf{SuccAKE_1}]$.

**Experiment ExpAKE₃**. We further modify the experiment as follows: The **ExpAKE₃** modification:

- The simulator chooses two random elements $A,B \in$ G and sets $Y = B$.

- For each fresh instance, the simulator chooses a random $r \in Z_q^*$ and sets $X = rA$. For other instances, the simulator computes $X$ as in experiment **ExpAKE₂**.

- For each fresh instance, the simulator sets each $k_{UG}$ to a random $\ell$-bit string. For other instances, the simulator computes $k_{UG}$ as in experiment **ExpAKE₂**.

Since $k_{UG}$ is set to a random $\ell$-bit string (for fresh instances), the success probability of $\mathcal{A}$ may be different between **ExpAKE$_2$** and **ExpAKE$_3$** if it makes an $J(T_U\|X\|Y\|K_{UG})$ query. This difference is bounded by Claim 7.

**Claim 7** | $\Pr_{P,A}[\mathsf{SuccAKE}_3] - \Pr_{P,A}[\mathsf{SuccAKE}_2] \mid\, \leq 1/q_J \cdot \mathsf{Adv}_G^{\mathrm{ECCDH}}(t_{AKE}^3)$, *where $q_J$ is the number of queries made to the J oracle.*

*Proof.* We prove the claim via a reduction from the ECCDH problem which is believed to be hard. Assume that the success probability of $\mathcal{A}$ is non-negligibly different between **ExpAKE$_2$** and **ExpAKE$_3$**. Then we construct an algorithm $\mathcal{A}_{\mathrm{ECCDH}}$ that solves the ECCDH problem in G with a non-negligible advantage. The objective of $\mathcal{A}_{\mathrm{ECCDH}}$ is to compute and output the value $W = uvP \in \mathrm{G}$ when given an ECCDH-problem instance $(U = uP, V = vP) \in \mathrm{G}^2$. $\mathcal{A}_{\mathrm{ECCDH}}$ runs $\mathcal{A}$ as a subroutine while simulating all the oracles on its own.

$\mathcal{A}_{\mathrm{ECCDH}}$ handles all the oracle queries of $\mathcal{A}$ as specified in experiment **ExpAKE$_3$** but using $U$ and $V$ in place of $X$ and $Y$. When $\mathcal{A}$ outputs its guess $b'$, $\mathcal{A}_{\mathrm{ECCDH}}$ chooses an entry of the form $(T_U\|X\|Y\|K_j)$ at random from JList and terminates outputting $K/r$. From the simulation, it is clear that $\mathcal{A}_{\mathrm{ECCDH}}$ outputs the desired result $W = uvP$ with probability at least $1/q_J$ if $\mathcal{A}$ makes a $J(T_U\|X\|Y\|K_{UG})$ query for some fresh instance of any $U \in \mathrm{U}$. This completes the proof of Claim 7.

**Experiment ExpAKE$_4$.** This experiment is different from **ExpAKE$_3$** in that it is aborted if the following event Forge occurs.

Forge: The event that the adversary $\mathcal{A}$ makes a Send query that contains a MAC forgery. Then we claim that:

**Claim 8** | $\Pr_{P,A}[\mathsf{SuccAKE}_4] - \Pr_{P,A}[\mathsf{SuccAKE}_3] \mid\, \leq q_{\mathrm{send}} \cdot \mathsf{Adv}_\Sigma^{\mathrm{EF-CMA}}(t_{AKE}^4)$, *where $q_{\mathrm{send}}$ is the number of queries made to the Send oracle.*

*Proof.* Assume that the event Forge occurs with a non-negligible probability. Then, we construct an algorithm $A_{\mathrm{EF}}$ who generates, with a non-negligible probability, a forgery against the MAC scheme $\Sigma$. The algorithm $A_{\mathrm{EF}}$ is given access to the $\mathsf{Mac}_k(\cdot)$ and $\mathsf{Ver}_k(\cdot)$ oracles. The goal of $A_{\mathrm{EF}}$ is to produce a message/MAC pair $(m,\sigma)$ such that: (1) $\mathsf{Ver}_k(m, \sigma) = 1$ and (2) $\sigma$ has not been output by the oracle $\mathsf{Mac}_k(\cdot)$ on input $m$.

Let $n$ be the total number of MAC keys used in the sessions initiated via a Send query. $A_{\mathrm{EF}}$ begins by choosing a random $i \in \{1, \ldots, n\}$. Let $k_i$ denote the $i^{\mathrm{th}}$ key among all the $n$ MAC keys, and Send$_i$ be any Send query that is expected to be answered and/or verified using $k_i$. $A_{\mathrm{EF}}$ runs $\mathcal{A}$ as a subroutine and answers the oracle queries of $\mathcal{A}$ as in experiment **ExpAKE$_3$** except that: it answers all Send$_i$ queries by accessing its $\mathsf{Mac}_k(\cdot)$ and $\mathsf{Ver}_k(\cdot)$ oracles. As a result, the $i^{\mathrm{th}}$ MAC key $k_i$ is not used during the simulation. If Forge occurs against an instance who holds $A_{\mathrm{EF}}$ halts and outputs the message/MAC pair generated by $\mathcal{A}$ as its forgery. Otherwise, $A_{\mathrm{EF}}$ terminates with a failure indication.

If the guess $i$ is correct, then the simulation is perfect and $A_{\mathrm{EF}}$ achieves its goal. Namely, $\mathsf{Adv}_\Sigma^{\mathrm{EF-CMA}}(A_{\mathrm{EF}}) = \Pr[\mathsf{Forge}]/n$. Since $n \leq q_{\mathrm{send}}$ and $A_{\mathrm{EF}}$ runs in time at most $t_{AKE}^4$, we get

$$\Pr[\mathsf{Forge}] \leq\ q_{\mathrm{send}} \cdot \mathsf{Adv}_\Sigma^{\mathrm{EF-CMA}}(A_{\mathrm{EF}})$$
$$\leq\ q_{\mathrm{send}} \cdot \mathsf{Adv}_\Sigma^{\mathrm{EF-CMA}}(t_{AKE}^4).$$

This completes the proof of Claim 8.

**Experiment ExpAKE$_5$.** We next modify the way of answering queries to the $H$ oracle as follows:Simulation of the $H$ oracle: For each $H$ query on a string $str$, the simulator first checks if an entry of the form $(str, h)$ is in a list called HList which is maintained to store input-output pairs of $H$. If it is, $h$ is the answer to the hash query. Otherwise, the simulator chooses a random $\kappa$-bit string $h'$, answers the query with $h'$, and adds the entry $(str, h')$ to HList.

The other oracle queries of $\mathcal{A}$ are handled as in experiment **ExpAKE$_4$**. Since **ExpAKE$_5$** is perfectly indistinguishable from **ExpAKE$_4$**, it is clear that:

**Claim 9.** $\mathrm{Pr}_{P,\mathcal{A}}[\mathsf{SuccAKE}_5] = \mathrm{Pr}_{P,\mathcal{A}}[\mathsf{SuccAKE}_4]$

**Experiment ExpAKE$_6$.** We finally modify the experiment so that the session key $sk$ is set to a random $\kappa$-bit string for each fresh instance and its partner. Accordingly, the success probability of $\mathcal{A}$ may be different between **ExpAKE$_6$** and **ExpAKE$_5$** if it asks an $H$ query of the form $H$ $(k_{US}\|T_U\|ID_{SN})$ for some uncorrupted $U \in \mathsf{U}$ and $SN \in \mathsf{SN}$. But the difference is bounded by:

**Claim 10** $\mid \mathrm{Pr}_{P,\mathcal{A}}[\mathsf{SuccAKE}_6] - \mathrm{Pr}_{P,\mathcal{A}}[\mathsf{SuccAKE}_5] \mid \leq \frac{1}{q_H} \cdot \mathsf{Adv}_{\Delta}^{\mathrm{IND-MEK}}(t_{AKE}^6)$, *where $q_H$ is the number of queries made to the $H$ oracle.*

*Proof.* We prove the claim by constructing an eavesdropper $\mathcal{A}_{\mathrm{IND\text{-}MEK}}$ who attacks the indistinguishability of $\Delta$ in experiment $\mathbf{Exp}_{\Delta}^{\mathrm{IND-MEK}}(A, n, d, b)$. $\mathcal{A}_{\mathrm{IND\text{-}MEK}}$ invokes the adversary $\mathcal{A}$ and answers all the oracle queries of $\mathcal{A}$ as in experiment **ExpAKE$_5$** except that it generates each $C_{GW}$ to be sent to a fresh sensor instance by accessing its own encryption oracle as follows:

Let $k'_{US} \neq k_{US}$ be a random string chosen from $\{0, 1\}^{\kappa}$. $\mathcal{A}_{\mathrm{IND\text{-}MEK}}$ outputs $(k_{US}, k'_{US})$ as a plaintext pair in the indistinguishability experiment $\mathbf{Exp}_{\Delta}^{\mathrm{IND-MEK}}$. Let $c$ be the ciphertext received in return for the plaintext pair. $\mathcal{A}_{\mathrm{IND\text{-}MEK}}$ sets $C_{GW}$ equal to the ciphertext $c$.

That is, each $C_{GW}$ is set to the encryption of either $k_{US}$ or $k'_{US}$. Now when $\mathcal{A}$ terminates and outputs its guess $b'$, $\mathcal{A}_{\mathrm{IND\text{-}MEK}}$ selects an entry of the form $(k_{US}\|T_U\|ID_{SN}, h)$ at random from HList and outputs 0 if $k = k_{US}$, and 1 otherwise. If $\mathcal{A}$ asks an $H$ query of the form $H$ $(k_{US}\|T_U\|ID_{SN})$ for some uncorrupted $U \in \mathsf{U}$ and $SN \in \mathsf{SN}$, $\mathcal{A}_{\mathrm{IND\text{-}MEK}}$ correctly guesses the bit $b$ in its indistinguishability experiment with probability at least $\frac{1}{q_H}$ and therefore, Claim 10 follows.

In experiment **ExpAKE$_6$**, the adversary $\mathcal{A}$ obtains no information on the hidden bit $b$ chosen by the $\mathsf{TestUA}$ oracle since the session keys of all fresh instances are selected uniformly at random from $\{0, 1\}^{\kappa}$. Therefore, it follows that $\mathrm{Pr}_{P,\mathcal{A}}[\mathsf{SuccUA}_4] = 1/2$. This result combined with Claims 5–10 completes the proof of Theorem 2.

## Concluding Remarks

With the continuing advancements in sensor technologies, WSNs will play an increasingly important role in commercial, government and military settings. A number of recent high profiles such as the revelations by Edward Snowden that the US National Security Agency has been conducting massive online surveillance of both US and non-US citizens highlighted the potential of ensuring user privacy and anonymity. In WSNs, for example, designing a secure and efficient user authentication scheme without compromising user anonymity remains an area of active research.

In this work, we have presented a SCA-WSN scheme, a smart-card-based user authentication scheme for wireless sensor networks, which achieves user anonymity without imposing (expensive) public key operations on sensors. Our result in this paper does not contradict the result of Wang and Wang [28, 29] but rather supports and clarifies it: *in order for a SCA-WSN scheme to achieve user anonymity, the use of public key cryptography is inevitable but, if forward secrecy is not desired, can be avoided at least on the sensor side.* Extending our result to the case of three-factor authentication [34] would be an interesting future work.

## Acknowledgments

All authors, especially the corresponding author Sangchul Han, would like to thank the anonymous reviewers for their time and invaluable comments and suggestions on this paper.

## Author Contributions

Conceived and designed the experiments: JN KKRC JP DW. Performed the experiments: SH JP MK. Analyzed the data: SH JP DW. Contributed reagents/materials/analysis tools: JP DW. Wrote the paper: JN KKRC SH MK JP DW. Designed the scheme: JN KKRC DW. Proved the security of the scheme: JN KKRC.